\documentclass[american, aps,pra, twocolumn, showpacs,reprint ]{revtex4-1} 
\usepackage[unicode=true,pdfusetitle, bookmarks=true,bookmarksnumbered=false,bookmarksopen=false, breaklinks=false,pdfborder={0 0 0},backref=false,colorlinks=false] {hyperref}
\hypersetup{ colorlinks,linkcolor=myurlcolor,citecolor=myurlcolor,urlcolor=myurlcolor}
\usepackage{braket,colortbl,cleveref,amsthm,amsmath,amssymb,txfonts}
\definecolor{myurlcolor}{rgb}{0,0,0.7}
\usepackage{graphics,graphicx}
\usepackage{color}
\usepackage{graphicx}
\usepackage[format = plain,labelfont = bf,up, textfont = normal , up, justification =raggedright, singlelinecheck =false]{caption}
\usepackage{subcaption}
\usepackage{tabularx}
\usepackage{amsmath}

\usepackage{graphicx}
\usepackage[utf8x]{inputenc}
\usepackage{color}
\usepackage{amsmath}
\usepackage{braket}
\usepackage{latexsym}
\usepackage{amssymb}
\usepackage{amsthm}
\usepackage{bm}
\usepackage{graphics,epstopdf}
\usepackage{color}\usepackage{amsmath}
\usepackage{enumitem}
\usepackage{fmtcount}
\usepackage{booktabs}

\usepackage{epsfig}

\theoremstyle{plain}

\def\bea{\begin{eqnarray}}
\def\eea{\end{eqnarray}}
\def\ba{\begin{array}}
\def\ea{\end{array}}

\def\ket{\rangle}
\def\bra{\langle}
\def\beq{\begin{equation}}
\def\eeq{\end{equation}}

\begin{document}

\title{A note on robustness of coherence for multipartite quantum states}
\author{Chiranjib Mukhopadhyay}
\affiliation{Harish-Chandra Research Institute$,$ Allahabad$, $India,\\ Homi Bhabha National Institute, Training School Complex, Anushakti Nagar, Mumbai 400 085, India}

\author{Udit Kamal Sharma, Indranil Chakrabarty}
\affiliation{Center for Security, Theory and Algorithmic Research, International Institute of Information Technology, Gachibowli, Hyderabad 500 032, India}

\begin{abstract}
\noindent{In this brief report, we prove that robustness of coherence (ROC), in contrast to many popular quantitative measures of quantum coherence derived from the resource theoretic framework of coherence, may be sub-additive for a specific class of multipartite quantum states. We investigate how the sub-additivity is affected by admixture with other classes of states for which ROC is super-additive. We show that pairs of quantum states may have different orderings with respect to relative entropy of coherence, $l_{1}$-norm of coherence and ROC and numerically study the difference in ordering for coherence measures chosen pairwise.}
\end{abstract}

\maketitle

\section{Introduction}

\noindent{Quantum Mechanics is the theoretical cornerstone underpinning our understanding of the natural world. The abstract laws of quantum mechanics also present us with resources we can harness to perform practical and important information theoretic tasks \citep{nielsen}. Motivated by the importance of quantum entanglement \citep{entanglement} in quantum communication schemes, a general study of the theory of resources within the quantum framework and beyond is being developed at present. One such concrete example of a quantum resource theory is the resource theory of coherence \citep{aberg,baumgratz,winter, manabrecent,adessorev, gour, broadcasting, plenio2, noc}, which seeks to quantify and study the amount of linear superposition a quantum state possesses with respect to a given basis. Since the superposition principle differentiates quantum mechanics from classical particle mechanics, quantum coherence may be viewed as a fundamental signature of nonclassicality in physical systems. Coherence may be considered a resource for certain tasks like better cooling \citep{huber, brask} or work extraction \citep{lostaglio1} in nanoscale thermodynamics, quantum algorithms \citep{algo1,algo2,algo3} or biological processes \citep{bio1,bio2,bio3}. The relation of resource theory of quantum coherence to resource theories of entanglement \citep{uttament,ent1,ent2,ent3,ent4,ent5,ent6,ent7} and thermodynamics \citep{lostaglio1, lostaglio2} is also quite close.

\noindent{However, any resource, including quantum coherence, may decay. One can thus quantify quantum  resources in terms of how robust they are against mixing with other states. This quantitative measure, introduced in literature \citep{roc,roa} as the \emph{Robustness of Coherence (ROC)} follows all the necessary and desirable conditions for a measure of quantum coherence laid down in \citep{baumgratz}. In this paper, we point out a surprising property of robustness of coherence. Unlike many other measures of quantum coherence, including two most popular measures viz. \emph{$l_{1}$-norm of coherence} and \textit{ relative entropy of coherence}, we show that robustness of coherence is not super-additive for multipartite quantum states in general. To this end, we explicitly point out a specific class of quantum states for every member of which, robustness of coherence of the multipartite state is less than the robustness of coherence of the sum of the reduced states. However, it is worth pointing out that for many classes of multipartite states, e.g. pure states or X-states, ROC is still super-additive. Thus, it is important to study how the superadditivity of quantum coherence gets affected if states from two different classes, one satisfying subadditivity of ROC and the other satisfying superadditivity of ROC, are mixed. Rather interestingly, we numerically observe that when states from a class of quantum states satisfying superadditivity of ROC are mixed with states from this class of states satistying sub-additivity of ROC , provided the mixing weight of the superadditive class of states exceeds a certain value, ROC of every such resulting mixed state is super-additive.}  

\noindent{We also address the issue of non-unanimous ordering of pairs of quantum states with respect to different coherence measures. While the ROC is identical to the $l_{1}$ norm of coherence and quite different from the relative entropy of coherence in two dimensional systems, we note that as we increase the dimension of the quantum system, a randomly chosen pair of quantum states is more likely to have different ordering with respect to $l_{1}$norm and ROC rather than with respect to relative entropy of coherence and ROC. This is in spite of the fact \citep{roc,roa} that for many multidimensional families of quantum states like  pure states or X-states, the ROC is identical to the $l_{1}$ norm of coherence. We observe a similar behaviour for randomly chosen higher rank states with a given dimension.}  

\noindent{The paper is organized as follows. In section II, we briefly recall the basic structure of resource theory of coherence and the definition of ROC. In section III, we prove two results for quantum coherence on bipartite systems. In section IV, we study the possible sub-additivity of ROC. Section V deals with the discussion on ordering of quantum states with respect to different coherence measures. We conclude in Section VI.}

\section{Robustness of Coherence}

\noindent At first, we shall look into the criteria needed by a functional $C$ to qualify as a measure of coherence. In Baumgratz's framework \cite{baumgratz}, a functional $C$, mapping quantum states to a non-negative real numbers, must satisfy the following properties to qualify as a measure of quantum coherence: 
\begin{itemize}
\item \textbf{[C1]} Firstly, C should vanish on all incoherent states : $C(\rho) = 0$ , $\forall \rho \in \textit{I}$, where \textit{I} is the set of all incoherent states in the given basis.

\item \textbf{[C2]} Secondly, $C$ should not increase under incoherent operations, which can be of types A and B.

\begin{itemize}

\item \textbf{[C2a]} Under type A operations, we have monotonicity under incoherent completely positive and trace preserving maps, that is, $C(\rho) \geq C(\Phi_{ICPTP}(\rho)), \forall \Phi_{ICPTP}$.

\item \textbf{[C2b]} Under type B operations, we have monotonicity under selective measurements on average, that is, $C(\rho) \geq \sum_{n}p_{n}C(\rho_{n}), \forall \{ K_{n} \}$ such that $\sum_{n} K_{n}^\dagger K_{n} = \mathbb{I}$ and $K_{n}IK_{n}^\dagger  \subset I$, where $I$ is the set of all incoherent states in the given basis.
\end{itemize}

\item \textbf{[C3]} Moreover, we would ideally like to ensure that coherence can only decrease under mixing, which leads to our final condition: non-increasing under mixing of quantum states (convexity), that is, $\sum_{n}p_{n}C(\rho_{n}) \geq C(\sum_{n}p_{n}\rho_{n})$ for any set of states $\{ \rho_{n} \}$ and any $p_{n} \geq 0$ with $\sum_{n}p_{n} = 1$.

\end{itemize}

\noindent{Now, we recall the definition of Robustness of coherence which satisfies all the above criteria for a coherence monotone.}

\noindent\emph{\textbf{Robustness of coherence (ROC) }- Let $\mathcal{D}(\mathcal{C}^{d})$ be the convex set of density operators acting on a \textit{d}-dimensional Hilbert Space. Let $\mathcal{I} \subset \mathcal{D}(\mathcal{C}^{d})$ be the subset of incoherent states. Then, the robustness of coherence (ROC) of a state $\rho \in \mathcal{D}(\mathcal{C}^{d})$ is defined as: \begin{equation} \label{eq1}
C_{\mathcal{ROC}}(\rho) = \min_{\tau \in \mathcal{D}(\mathcal{C}^{d})}
\left \{ s \geq 0 \mid \frac{\rho + s\tau}{1 + s} = \colon \delta \in \mathcal{I}
\right \}.
\end{equation}}
Clearly  $C_{\mathcal{ROC}}(\rho)$ is the minimum weight of another state $\tau$ such that its convex mixture with $\rho$ yields an incoherent state $\delta$. It is slightly different from the similarly defined robustness of entanglement \citep{roe} in that the mixing is not only over free, i.e. incoherent states in this case.

\noindent{The robustness of coherence has an operational interpretation as a coherence witness through a semidefinite program. It also means that $C_{\mathcal{ROC}}(\rho)$ can be evaluated via a semidefinite program that finds the optimal coherence witness operator. This semidefinite program has been used to carry out the numerical calculations in this paper.}

\section{Preliminary Results}

In this section we derive two results on quantum coherence for joint states. 
 \\
\noindent\textit{Result I: For any pure state $|\psi_{AB}\rangle$, ROC is super-additive.}  
\begin{proof} 
For any pure state $|\psi_{AB}\rangle$, we have $C_{ROC}(|\psi_{AB} \rangle) = C_{l_{1}}(|\psi_{AB} \rangle)$. Now, we use the superadditivity of $l_{1}$-norm of coherence along with the fact that  ROC is always upper bounded by the the $l_{1}$-norm of coherence to obtain
$C_{ROC}(|\psi_{AB} \rangle) = C_{l_{1}}(|\psi_{AB} \rangle) \geq  C_{l_{1}}(\rho_{A}) + C_{l_{1}}(\rho_{B} ) \geq C_{ROC}(\rho_{A}) + C_{ROC}(\rho_{B})$, thus proving the result.
\end{proof}
We now show that adding an incoherent ancilla doesn't change the amount of coherence in a system.  This   intuitively obvious statement is shown below to hold for arbitrary legitimate coherence measures. In order to prove this, we note that the  inequality \textbf{[C2b]} has been shown as equivalent \citep{c2balt} to the equality condition that if $\rho = p_{1} \rho_{1} \oplus p_{2} \rho_{2}$ for $p_{1} + p_{2} = 1$, then
\begin{equation}
\label{C3alternate}
C(p_{1} \rho_{1} \oplus p_{2} \rho_{2}) = p_{1} C(\rho_{1}) + p_{2} C(\rho_{2}).
\end{equation}

\noindent \textit{Result II: For any state $\rho_{A}$ and any incoherent state $\sigma_{B}$,   $C(\rho_{A} \otimes \sigma_{B}) = C(\rho_{A})$ for any legitimate coherence measure $C$.}

\begin{proof}
\noindent{Let us assume that $dim(\mathcal{H}_{A}) = dim(\mathcal{H}_{B})= n \geq 2$. Clearly, $\rho_{A} \otimes \sigma_{B}$ is a $n^{2}$ x $n^{2}$ sparse matrix with its sparsity = $1 - \frac{1}{n} \geq \frac{1}{2}$. As the dimension $n$ increases, the sparsity also increases. Let $X = \rho_{A} \otimes \sigma_{B}$. Given a sparse matrix, we can always use permutation matrices to transform it to a matrix in block-diagonal form \citep{tewarson} $ =d_{1} \rho_{A} \oplus d_{2} \rho_{A} \oplus .... \oplus d_{n} \rho_{A}$ via permutation matrices \footnote{Since permutations correspond merely to relabeling of the basis vectors, amount of coherence of a system does not depend on such permutations}. Now, from \eqref{C3alternate} \citep{c2balt}, we have, for any legitimate coherence measure $C$, $C(\rho_{A}\otimes \sigma_{B}) = C (d_{1} \rho_{A} \oplus d_{2} \rho_{A} \oplus .... \oplus d_{n} \rho_{A}) = \sum_{i=1}^{n} d_{i} C(\rho_{A}) = C(\rho_{A})$, where the last line follows from the unit trace condition for density matrices.}
 
\end{proof}

\section{Sub-additivity of Robustness of Coherence}

\noindent{In this section, we explore the possible sub-additivity of ROC. To this end, we introduce the following class of $n$-qubit states $\rho_{A_{1}A_{2}A_{3}....A_{n}} = (1+k) \frac{\mathbb{I}}{2^{n}} - k |\psi \rangle \langle \psi|$, where $\mathbb{I}$ is the identity matrix, $0 \leq k \leq \frac{1}{2^{n} - 1}$ and $|\psi \rangle = \frac{1}{2^{n/2}}(\sum_{i=1}^{2^{n}}|i\rangle)$ is the maximally coherent $n$-qubit state.}\\

 
\textit{Theorem I - For an arbitrary n qubit system $A_{1}A_{2}A_{3}....A_{n}$, the ROC for the family $\Sigma$ of states $ \rho_{A_{1}A_{2}A_{3}....A_{n}} = (1+k) \frac{\mathbb{I}}{2^{n}} - k |\psi \rangle \langle \psi|$, where $0 \leq k \leq \frac{1}{2^{n} - 1}$ and $|\psi \rangle = \frac{1}{2^{n/2}}(\sum_{i=1}^{2^{n}}|i\rangle)$ is the maximally coherent n-qubit state, satisfies the following sub-additive relation :}
 \begin{equation}
 \label{subadditivity}
 C_{ROC} (\rho_{A_{1}A_{2}A_{3}....A_{n}}) \leq \sum_{i = 1}^{n} C_{ROC} (\rho_{A_{i}}). 
 \end{equation}

\begin{proof}
\noindent{Given, $\rho_{A_{1}A_{2}A_{3}....A_{n}} = (1+k) \frac{\mathbb{I}}{2^{n}} - k |\psi \rangle \langle \psi|$, where $0 \leq k \leq \frac{1}{2^{n} - 1}$ and $|\psi \rangle$ is the maximally coherent n-qubit state. Now, by using definition of robustness of coherence \eqref{eq1}, we prepare a convex mixture $\chi$ of an arbitrary n-qubit state $\tau$ and $\rho_{A_{1}A_{2}A_{3}....A_{n}}$, that is, mathematically expressed as :}
 \begin{equation}
 \label{X}
\chi = \frac{(1+k) \frac{\mathbb{I}}{2^{n}} - k |\psi \rangle \langle \psi| + s\tau}{1 + s},
 \end{equation}
\noindent{where $s$ is $C_{ROC} (\rho_{A_{1}A_{2}A_{3}....A_{n}})$. Without any loss of generality, when $\chi$ in Eq.\eqref{X} is expanded in n-qubit computational basis, the diagonal elements are of the form 
\begin{equation}\label{diagonal}
 \chi_{ii} = \frac{1 + 2^{n}s\tau_{ii}}{2^{n}(1 + s)},
\end{equation}
whereas, the off-diagonal elements are of the form 
\begin{equation}\label{offdiagonal}
 \chi_{ij} = \frac{-k + 2^{n}s\tau_{ij}}{2^{n}(1 + s)}.
\end{equation}
}

\noindent{For $\chi$ in Eq.\eqref{X} to be an incoherent state, we have to ensure that the off-diagonal elements of $\chi$, described by Eq.\eqref{offdiagonal}, will be zero. So, by equating Eq.\eqref{offdiagonal} to zero, we finally arrive at the following condition :
\begin{equation}\label{condition1}
s =  \frac{k}{2^{n}\tau_{ij}}.
\end{equation}}

\noindent{As per definition of robustness of coherence(Eq. \eqref{eq1}), $s \in \Re$, where $\Re$ is the set of Real numbers, has to be minimized. Since $s \in \Re$, so, clearly,$\tau_{ij} \in \Re$. Now, in the trivial case, $s$ is zero when $\rho_{A_{1}A_{2}A_{3}....A_{n}}$ is already an incoherent state. In the non-trivial case, $s$ is minimum when $\tau_{ij}$ takes the maximum value of $k$, i.e, $\tau_{ij} = \frac{1}{2^{n} - 1}$. Hence, after substituting $\tau_{ij} = \frac{1}{2^{n} - 1}$ in Eq.\eqref{condition1}, we have,}

\begin{equation}\label{CRhoBig}
s = C_{ROC} (\rho_{A_{1}A_{2}A_{3}....A_{n}})
  = k\bigg( 1 - \frac{1}{2^{n}}\bigg).
\end{equation}

\noindent{Now, let us consider the single-qubit subsystems $\rho_{A_{i}} = Tr_{A_{1}....A{i - 1}A{i + 1}...A{n}} [\rho_{A_{1}A_{2}A_{3}....A_{n}}] = \left(\begin{matrix}
\frac{1}{2} & -\frac{k}{2}\\ -\frac{k}{2} & \frac{1}{2}
\end{matrix} \right)$ in computational basis. For single qubit systems, we know that robustness of coherence is equal to its $l_{1}$-norm of coherence for a fixed basis. Hence, for single qubit computational basis, $C_{ROC}(\rho_{A_{i}})$ = $C_{l_{1}}(\rho_{A_{i}})$ = $k$.}\\

\noindent{Finally, we have, \begin{equation} \label{final}
\begin{split}
\Lambda &= C_{ROC} (\rho_{A_{1}A_{2}A_{3}....A_{n}}) - \sum_{i = 1}^{n} C_{ROC} (\rho_{A_{i}})\\
&= k\bigg( 1 - \frac{1}{2^{n}}\bigg) - nk \\
&= k\bigg[ 1 - \bigg(n + \frac{1}{2^{n}}\bigg) \bigg].
\end{split}
\end{equation}
}\\

\noindent{Clearly, for $n \in \mathbb{Z^{+}}$ and $0 \leq k \leq \frac{1}{2^{n} - 1}$, we have $\Lambda \leq 0$. Hence, proved.}\\

\end{proof}

\noindent{Given any pure state, its ROC is identical with its $l_{1}$-norm of coherence, which is always super-additive. We now turn to the scenario when elements of the set of states $\Sigma$ mentioned in the previous theorem are mixed with a given pure state $|\phi \rangle$ and investigate what happens to the  sub-additivity property as we increase the mixing.
}\\

To this end, we randomly pick a large number of states $\lbrace \sigma \rbrace$ from the family $\Sigma$ and mix every such  state with a chosen pure state $|\phi\rangle$ with mixing parameter $p$ to obtain a large number of states $\Sigma_{p} = \lbrace (1-p)\sigma + p |\phi \ket \bra \phi| \rbrace$. We want to know the probability of any randomly chosen element of this set satisfying the sub-additivity condition Eq.\eqref{subadditivity}. Clearly, if $p = 0$, this set is a random subset of $\Sigma$, thus all the elements will satisfy the subadditivity  condition. In the opposite limit, if $p =1$, this set consists of only $|\phi\rangle$, i.e. always super-additive for ROC. However, it is the intermediate region which is of interest to us. For simplicity, we confine ourselves to the 2-qubit scenario. We consider two different pure states $\phi\rangle$, one being the maximally coherent state $|\phi_{1}\rangle = \frac{1}{2} \left(|00\rangle + |01\rangle +|10\rangle + |11\rangle \right)$, the other being the maximally entangled state $|\phi_{2}\rangle = \frac{1}{2} \left(|00\rangle + |11\rangle \right)$. For each of them and every value of the mixing weight $p$, choosing 10,000 random states from $\Sigma_{p}$ according to Haar measure, we calculate the percentage of states in the set $\Sigma_{p}$ which satisfy the subadditivity condition. FIG \ref{subAdd} captures the result. Two properties of this figure are quite interesting. Firstly, the plots are almost identical for two very different sets of pure states $|\phi_{i} \rangle [i = 1,2] $, viz. the maximally coherent states(indicated by red points) and the maximally entangled states(indicated by blue points). Secondly, instead of the proportion of states satisfying subadditivity condition \eqref{subadditivity}    diminishing smoothly as $p \rightarrow 1 $, it shows a sudden death at around $p = 0.25$.

\begin{figure}[htbp]
\begin{center}

\includegraphics[height=6cm,width=6cm]{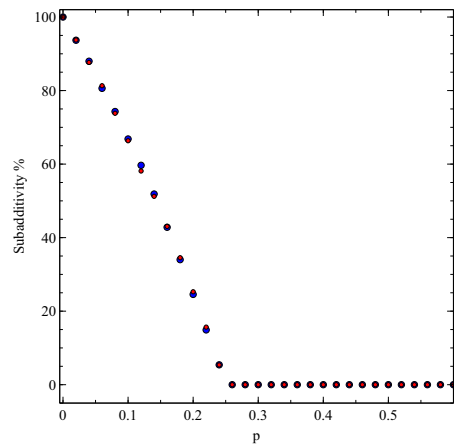}\\

\end{center}
\caption{(color online) Percentage of randomly chosen two qubit states from $\Sigma_{p}$ which satisfies subadditivity vs. the mixing weight p, where the pure state $|\phi\rangle$ is either the two qubit maximally coherent state (red dots) or the two qubit maximally entangled state(blue dots). 1000 randomly generated states taken for rach value of p.}
\label{subAdd}
\end{figure}

\section{Ordering of states through different coherence measures}

\noindent{Quantification of any resource through some measure begs the question - what is the operational significance of that particular measure ? Indeed the same resource can be operationally relevant in many different protocols. This naturally leads us to the next question: if the same resource is quantified by  different measures motivated by  different protocols - then can a state which is \emph{bad} for a particular protocol turn out to be \emph{good} for another protocol utilizing the same resource ?
}\\

\noindent{For the resource theory of coherence - a central question is, when can one transform a quantum state $\rho$ to $\sigma$ using incoherent operations ? If both input and target states are pure, say $ |\psi \rangle$ and $ |\chi \rangle$ respectively , a necessary and sufficient condition for such convertibility  \citep{icopure} is given by: 
\begin{equation}
\label{pure_convertibility}
\vec{c}_{\psi} \prec \vec{c}_{\chi},
\end{equation} 
where $\vec{c}_{\xi}$ for any state $|\xi \rangle$ is the collection of squared moduli of the coefficients of that state when expanded out in the basis of our choice. Evidently it is possible to have pairs of pure states for which the collection of coefficients  do not majorize each other. This leaves open the possibility that even for pairs of such pure states, two different coherence measures may give us different ordering. This is indeed confirmed for pure as well as mixed states  \citep{ordering_violation} for $C_{rel}$ and $C_{l_{1}}$. In this section, we investigate the statistics of  ordering for different coherence measures, viz. $C_{rel}$, $C_{l_{1}}$ and $C_{ROC}$, if random states are chosen from the state space according to Haar measure. We decided to check the percentage of randomly chosen pairs of states with different ordering wrt pairwise chosen coherence measures depending upon dimension and rank of the chosen states. \footnote{Why both dimension and rank ? The explanation is that for higher dimensional states, when we generate datasets of random quantum states, we miss out the states of lower ranks which are of measure zero.}

\noindent{From Figure \ref{Violation1}, it is evident that as the dimension of the quantum state increases, the percentage of ordering violations between robustness of coherence and relative entropy measure of coherence (denoted by the green curve) remains greater than that of between robustness of coherence and $l_{1}$-norm of coherence (denoted by the blue curve) and $l_{1}$-norm and relative entropy of coherence (denoted by the red curve). Moreover, we observe that for dimension $d \leq 5$, the percentage of ordering violation between $l_{1}$-norm and relative entropy of coherence is greater than that of between robustness of coherence  and $l_{1}$-norm of coherence. However, for dimensions  $d > 5$, the percentage of ordering violation between $l_{1}$-norm and robustness of coherence is greater than that of between relative entropy of coherence and $l_{1}$-norm of coherence.}\\

\begin{figure}[htbp]
\begin{center}

\includegraphics[height=6.2cm,width=6.2cm]{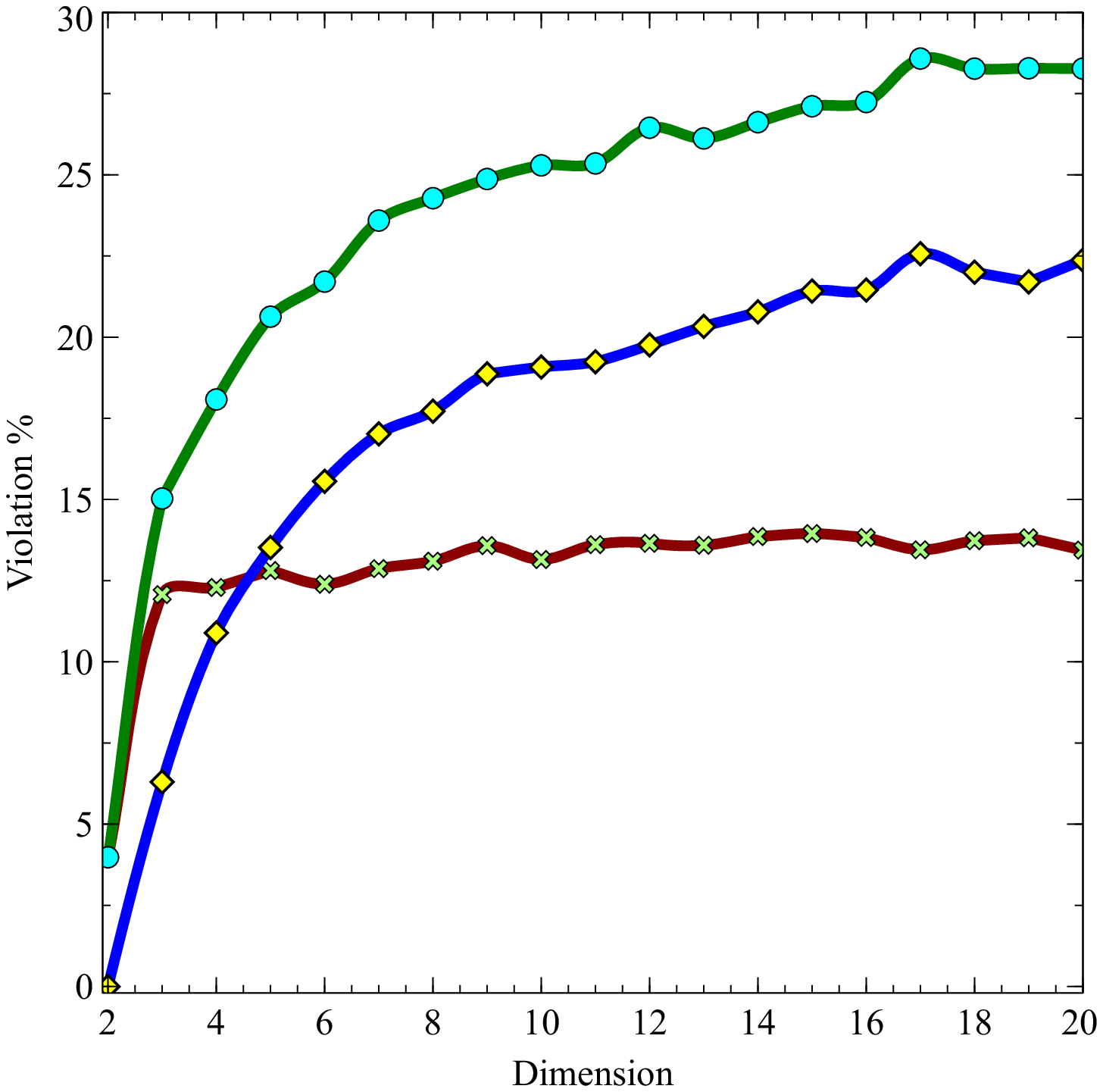}

\end{center}
\caption{(color online) Percentage of pairs of states with different ordering with respect to pairwise chosen coherence measures: $l_{1}$-norm vs relative entropy of coherence by the brown line, $l_{1}$-norm vs ROC by the blue line and relative entropy of coherence vs ROC  by the green line (taking $10,000$ randomly chosen pairs of  states) vs. dimension of states.  }
\label{Violation1}
\end{figure}

\noindent{In Figure \ref{Violation2}, we observe a similar trend as that of Figure \ref{Violation1}. Here, as the rank of the quantum state increases, the percentage of ordering violations between robustness of coherence and relative entropy measure of coherence is significantly greater than that of between robustness of coherence and $l_{1}$-norm of coherence  and $l_{1}$-norm and relative entropy of coherence. For pure states, i.e. states of rank 1, robustness of coherence is identical to the $l_{1}$ norm of coherence , therefore there is no ordering violation among them. However, for mixed states , the percentage of ordering violation between $l_{1}$-norm and robustness of coherence is greater than that of between relative entropy of coherence and $l_{1}$-norm of coherence.}\\

\begin{figure}[htbp]
\begin{center}

\includegraphics[height=6.5cm,width=6cm]{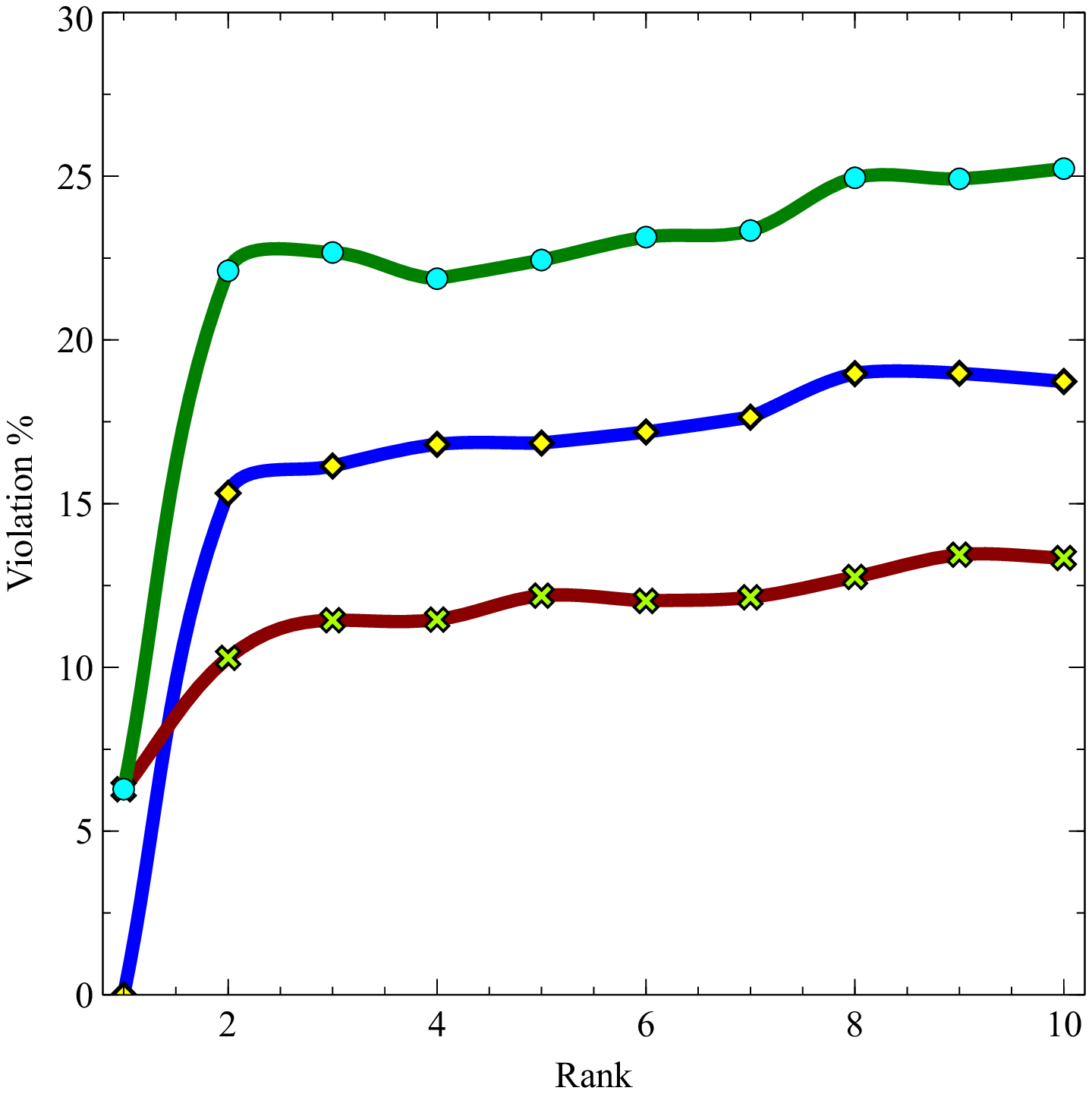}

\end{center}
\caption{(color online) Percentage of pairs of states with different ordering with respect to pairwise chosen coherence measures: $l_{1}$-norm vs relative entropy of coherence by the brown line, $l_{1}$-norm vs ROC by the blue line and relative entropy of coherence vs ROC  by the green line (taking $10,000$ randomly chosen pairs of  states) vs. rank of states (dimension d = 10).   }
\label{Violation2}
\end{figure}

\section{Conclusion}

We conclude that unlike $l_{1}$-norm or relative entropy of coherence, which are super-additive, ROC can be sub-additive for certain classes of states. If we take a mixture of that class of states and pure states,  we have found out that beyond a certain range of mixing weight, such mixtures cease to satisfy sub-additive property. We have found that for a pair of randomly generated density matrices, there exists a possibility of ordering violations corresponding to different legitimate measures of coherence. We welcome further work on implications of sub-additivity of ROC for quantum advantage in phase discrimination tasks and quantum information theory in general. 

\section{Acknowledgement}

CM acknowledges doctoral research fellowship from Department of Atomic Energy, Govt of India.  

\bibliographystyle{apsrev4-1}
\bibliography{roc}

\end{document}